\documentclass[manuscript]{aastex}

\usepackage{tikz}
\usepackage{amsmath}

\usepackage{graphicx}

\usepackage{pdflscape}

\slugcomment{\textbf{The Astronomical Journal} }

\shorttitle{Development of K2}
\shortauthors{Jewitt}

\begin{document}

\title{Distant Comet C/2017 K2  and the Cohesion Bottleneck}


\author{David Jewitt$^{1,2}$
Jessica Agarwal$^3$, Man-To Hui$^{1}$, Jing Li$^{1}$,  Max Mutchler$^4$, and \\ Harold Weaver$^5$
} 
\affil{$^1$Department of Earth, Planetary and Space Sciences,
UCLA, 
595 Charles Young Drive East, 
Los Angeles, CA 90095-1567\\
$^2$Department of Physics and Astronomy,
University of California at Los Angeles, \\
430 Portola Plaza, Box 951547,
Los Angeles, CA 90095-1547\\
$^3$ Max Planck Institute for Solar System Research, Justus-von-Liebig-Weg 3, 37077 G\"ottingen, Germany\\
$^4$ Space Telescope Science Institute, 3700 San Martin Drive, Baltimore, MD 21218 \\
$^5$ The Johns Hopkins University Applied Physics Laboratory, 11100 Johns Hopkins Road, Laurel, Maryland 20723  \\
}

\email{jewitt@ucla.edu}

\begin{abstract}
Distant long-period comet C/2017 K2 (PANSTARRS) has been outside the planetary region of the solar system for $\sim$3 Myr, negating the possibility that  heat retained from the previous perihelion could be responsible for its activity.  This inbound comet is also too cold for water ice to sublimate and too cold for amorphous water ice, if present, to crystallize.  C/2017 K2 thus presents an ideal target in which to investigate the mechanisms responsible for activity in distant comets.  We have used Hubble Space Telescope to study the comet in the pre-perihelion heliocentric distance range  13.8 $\le r_H \le$  15.9 AU.    In this range, the coma maintains a logarithmic surface brightness gradient $m = -1.010\pm$0.004, consistent with mass loss proceeding in steady-state.  The absence of a radiation pressure swept tail indicates that the effective particle size is large (radius $\gtrsim$ 0.1 mm) and the mass loss rate is $\sim$200 kg s$^{-1}$, remarkable for a comet still beyond the orbit of Saturn.  Extrapolation of the photometry indicates that activity began in 2012.1$\pm$0.5, at  $r_H$ = 25.9$\pm$0.9 AU, where the isothermal blackbody temperature is only $T_{BB}$ = 55 K.    
This large distance and low temperature suggest that cometary activity is driven by the sublimation of a super-volatile ice (e.g.~CO), presumably preserved by K2's long-term residence in the Oort cloud.    The mass loss rate can be sustained by CO sublimation from an area $\lesssim 2$ km$^2$, if located near the hot sub-solar point on the nucleus.
However, while the drag force from sublimated CO is sufficient to lift millimeter sized particles against the gravity of the cometary nucleus, it is 10$^2$ to 10$^3$ times too small to eject these particles against inter-particle cohesion.  Our observations thus require either a new understanding of the physics of inter-particle cohesion or the introduction of another mechanism to drive distant cometary mass loss.  We suggest thermal fracture and electrostatic supercharging in this context.

\end{abstract}


\section{INTRODUCTION}
\label{intro}
The sublimation of water ice can drive cometary activity out to about the orbit of Jupiter (5 AU), as proposed in the original ``dirty snowball'' model of the nucleus by Whipple (1950).  However, some comets, of which the long period comet C/2017 K2 (Pan STARRS) (hereafter ``K2'') is a prime  example, show activity at much larger distances.   Comet K2 was active at discovery (heliocentric distance $r_H$ = 15.9 AU) and was later found to be active in pre-discovery data out to $r_H$ = 23.7 AU (Jewitt et al.~2017, Meech et al.~2017, Hui et al.~2018).  Significantly, K2 is on a long-period orbit and approaching perihelion, which removes the possibility that the observed mass loss is driven by the activation of sub-surface volatiles by conducted heat acquired near perihelion.  Comet K2 is entering the planetary region of the solar system from Oort cloud distances and temperatures, retaining negligible heat from the previous perihelion.

The osculating orbit of K2 is marginally hyperbolic (semimajor axis $a$ = -5034 AU, perihelion distance $q$ = 1.810 AU, eccentricity $e$ = 1.00036 and inclination $i$ = 87.5\degr) with perihelion expected on UT 2022 December 21.   The slight excess eccentricity above unity  is a product of planetary perturbations, as shown by numerical integrations which reveal a pre-entry semi-major axis $a \sim$ 20,000 AU, eccentricity $e$ = 0.9998  and orbit period $\sim$ 3 Myr (Krolikowska and Dybczynski 2018).   K2  is thus  not dynamically new in the Oort sense, but is a return visitor  to the planetary region.  Crucially, however,  with the previous perihelion occurring $\sim$ 3 Myr ago, it is clear that current activity in K2 cannot be related  to  heat retained from the previous approach.  

The scientific excitement of K2 is that it provides an opportunity to study the rise of cometary activity in a never-before observed distance regime.   Hubble Space Telescope (HST) observations of K2 from a single orbit were reported in Jewitt et al.~(2017a).  In this paper, we describe continuing observations with HST, taken to study the early development of the coma as K2 approaches the Sun.  During our observations the comet approached from $r_H$ = 15.9 AU to 13.8 AU.  Our HST data have the  advantages of high angular resolution and of being taken with a single observational system.      

\section{OBSERVATIONS}
For the present observations we used images from the WFC3 imaging camera on the 2.4 m Hubble Space Telescope, taken  under observational programs GO 15409 and 15423.  The UVIS channel of this instrument uses two thinned, backside-illuminated charge-coupled devices (CCDs) each with 2051$\times$4096 pixels, separated by a 35 pixel gap.  The  image scale is 0.04\arcsec~pixel$^{-1}$ across a field 162\arcsec$\times$162\arcsec.  To reduce overhead times, we read out only a smaller CCD subarray, giving an 81\arcsec$\times$81\arcsec~field of view.  We used the wide F350LP filter in order to maximise sensitivity to faint sources in the data.  This filter benefits from a peak system throughput of 29\% and takes in most of the optical spectrum with wavelengths $\lambda >$ 3500\AA.  The effective wavelength on a sun-like spectrum source is 5846\AA~and the effective full-width at half maximum (FWHM) is  4758\AA.  A journal of observations is given in Table (\ref{geometry}), where dates are expressed as  Day of Year (DOY) with DOY = 1 on UT 2017 January 01.  

\begin{figure}
\epsscale{0.99}
\plotone{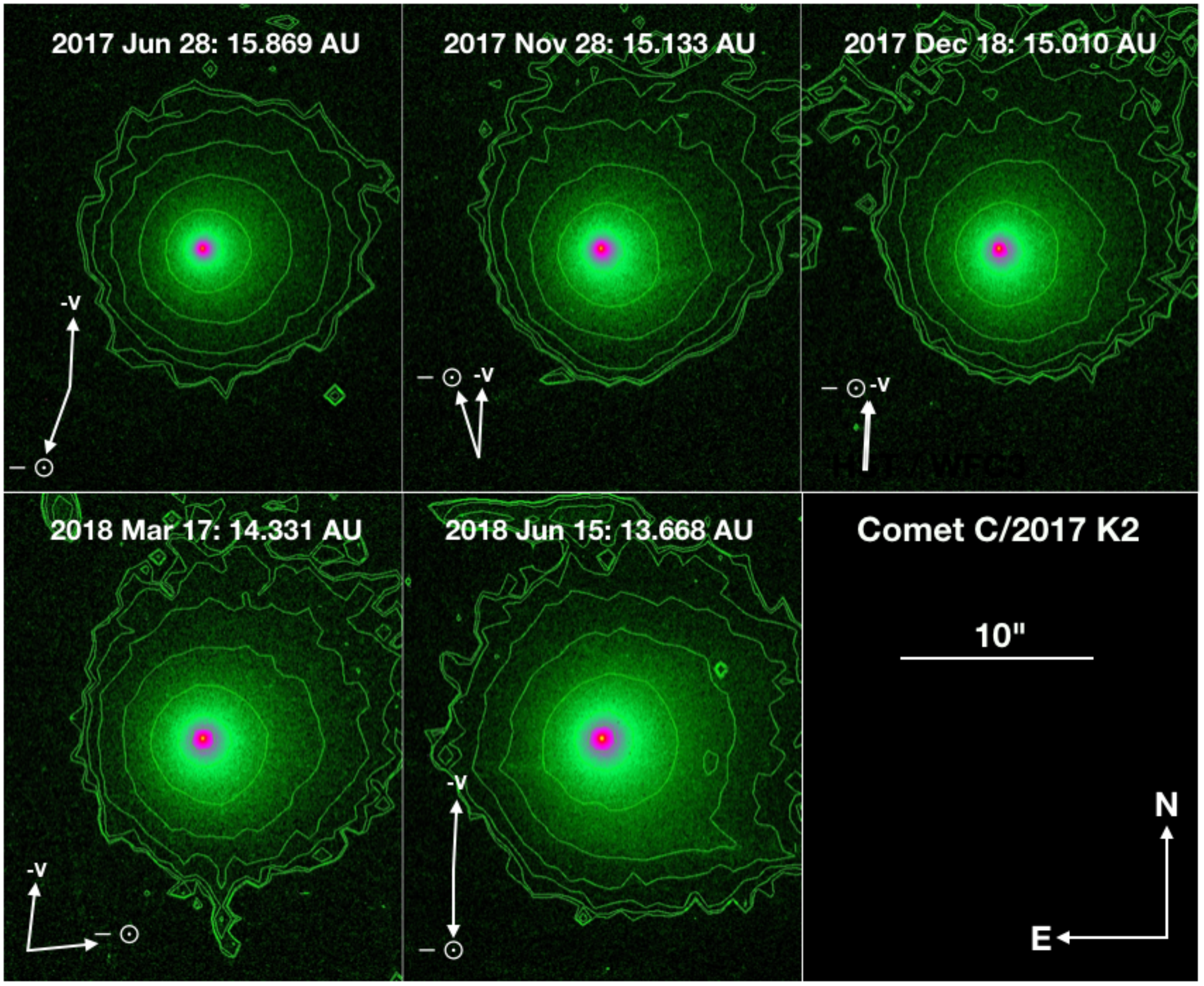}
\caption{Hubble images of C/2017 K2 marked with their UT dates (c.f.~Table \ref{geometry}).  White arrows show the projected negative heliocentric velocity vector, marked -V, and the anti-solar direction, marked $-\odot$, respectively.  Linear features (e.g.~2018 Mar 17 and Jun 15 panels) are the imperfectly removed trails of field stars and galaxies. A scale bar and the cardinal directions are marked at the lower right.  
\label{images}}
\end{figure}

In each  HST orbit we obtained six integrations of 285 s each (1710 s per orbit). The individual images from HST are strongly affected by cosmic rays.  A cosmetically clean image was obtained by  computing the median of the six images from each orbit after first shifting them to a common center.  The median images were also rotated to the correct orientation (north up and east left) prior to measurement (Figure \ref{images}).  The coma is distinguished by its largely circular isophotes, especially inside projected angles $\theta \le$ 5\arcsec, and by the absence of any clear change in the morphology as the Earth moved from one side of the orbital plane to the other (plane-crossing corresponds to the UT 2017 December 18 panel in Figure \ref{images}).

\subsection{Surface Brightness Profile}
We computed the surface brightness profile, $\Sigma(\theta)$, as a function of the angular distance, $\theta$, measured from  the photocenter of the comet.  For this purpose we used a set of 100 annular digital apertures, each 4 pixels (0.16\arcsec) wide and extending to 400 pixels (16\arcsec) radius.  We determined the sky background in an adjacent annulus extending from 400 to 500 pixels (16\arcsec~to 20\arcsec).  The limitation to the surface brightness measurements is largely set by the accuracy of sky background subtraction.  In HST data the background appears faintly structured,  above photon and detector noise, owing to field stars and galaxies trailed across the images by the non-sidereal tracking of the telescope that is required to keep K2 fixed (see Figure \ref{images}).  These trails are suppressed by computation of the orbital median image but typically cannot be entirely removed.  As a result, the sky uncertainty is non-gaussian and varies from image to image and orbit to orbit, depending on the background field.  We used image statistics within the sky annulus and inspection of line plots through the sky regions in order to estimate the sky uncertainty.

The main result is that the profile of K2 did not change in the 2017 - 2018 period.  In Figure (\ref{SB}) we show profiles from the first (UT 2017 June 28; green symbols) and last (UT 2018 June 15; yellow symbols) observations, together with their uncertainties.  In the central region $\Sigma(\le 0.2\arcsec)$ is affected by convolution of the intrinsic cometary profile with the point spread function of HST.  In the outer region, $\theta \gtrsim 2\arcsec$, the effects of sky subtraction uncertainty become significant.  In the 0.2\arcsec~$\le \theta \le$ 2.0\arcsec~ range, the surface brightness is well-fitted by a power-law, $\Sigma(\theta) \propto \theta^{m}$, with $m = -1.008\pm0.004$ on UT 2017 June 28 and $m = -1.012\pm0.004$ on UT 2018 June 15.  Within the uncertainties, these values are equivalent and  the mean, $m = -1.010\pm0.004$, is very close to the $m$ = -1 slope expected of a coma produced in  steady-state (Jewitt and Meech 1987).  Although the measurement on UT 2018 June 15 is formally different from $m$ = -1 at the 3$\sigma$ confidence level, we prefer to interpret the fits as setting a 3$\sigma$ limit to the gradient of $m > -1.012$.

The surface brightness profiles of K2 are remarkable for their stability, and for their closeness to the $m =$ -1 profile expected of a steady-state coma in the absence of radiation pressure.  Radiation pressure acceleration of dust particles should, in the limiting case, produce a tail and steepen the gradient measured within concentric circular apertures to $m$ = -3/2 (Jewitt and Meech 1987), very different from the measured gradient.   
The faintest isophotes in Figure (\ref{images}) do show asymmetry towards the north west that, if real, might be caused by solar radiation pressure, but deeper observations and better background removal are needed to establish the reality of the faint, outer regions.

\subsection{Aperture Photometry}

 On a diffuse source, the choice of photometry aperture is critical to the quantitative interpretation of the data.  We scaled the photometry apertures inversely with the instantaneous geocentric distance, $\Delta$, so as to maintain fixed linear radii of 5, 10, 20, 40, 80 and 160$\times$10$^3$ km when projected to  the distance of the comet.  The smallest of these apertures corresponds to an angular radius $\sim$0.4\arcsec~at geocentric distance $\Delta$ = 15 AU.  This is $\sim$10$\times$ the full width at half maximum of the point spread function, sufficient to avoid complications in the interpretation of the photometry caused by the finite resolution of the data.  The background sky brightness and its uncertainty were estimated from the median and dispersion of data numbers in a concentric annulus with inner and outer radii of 400 and 500 pixels (16\arcsec~and 20\arcsec), respectively.  The photometry was calibrated assuming that a G2V source with V = 0 would give a count rate 4.72$\times$10$^{10}$ s$^{-1}$ in the same filter.    In the interests of uniformity, we also re-analysed the images from the preceding program GO 14939 (c.f.~Jewitt et al.~2017a) using exactly the same procedures as applied to the new data.  Results from the re-analysis are consistent with those reported earlier, giving confidence that no systematic effects exist between the two datasets.

The photometric measurements are listed in Table (\ref{photometry}) where, in addition to the apparent magnitudes the Table lists the absolute magnitudes, $H$, and the effective scattering cross-sections of the comet, $C_e$, computed as follows.  The absolute magnitude, $H$, is the magnitude that would be observed if the comet could be relocated to unit heliocentric and geocentric distances, $r_H = \Delta$ = 1 AU, and to phase angle $\alpha$ = 0\degr.  We computed $H$ from the inverse square law, expressed as

\begin{equation}
H = V - 5\log_{10}(r_h \Delta) - f(\alpha)
\end{equation}

\noindent where $V$ is the apparent magnitude and $f(\alpha)$ is the phase function.   The phase functions of comet dust are in general poorly known and, in K2, the phase function is unmeasured. However, K2 was observed at very small and barely changing phase angles (Table \ref{geometry}) so that phase corrections are not a significant source of error.  We used $f(\alpha) = 0.04\alpha$, but values half or twice as large would introduce insignificant relative errors in the derived $H$ magnitudes.

The absolute magnitude is related to the effective scattering cross-section, $C_e$ [km$^2$], by 
\begin{equation}
C_e = \frac{1.5\times 10^6}{p_V} 10^{-0.4 H}
\label{area}
\end{equation}

\noindent where $p_V$ is the geometric albedo.  We assume $p_V$ = 0.04, consistent with the low albedos measured for the surfaces of cometary nuclei.  The apparent and absolute magnitudes and the scattering cross-sections are listed  in Table (\ref{photometry}) for each aperture and date of observation using the compact format $V/H/C_e$.

\section{DISCUSSION}

\subsection{Initiation of the Activity}
\label{production}
\begin{figure}
\epsscale{0.8}
\plotone{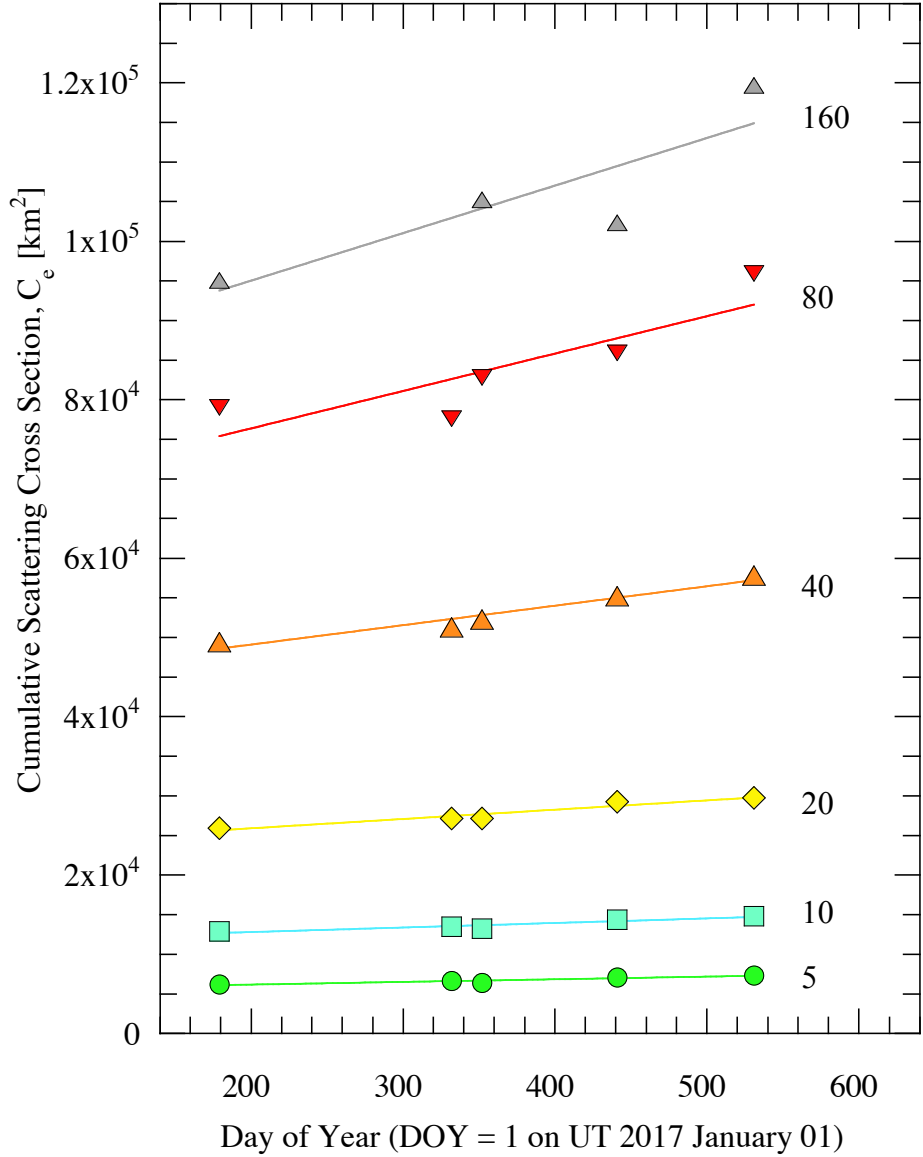}
\caption{Scattering cross-section as a function of the date of observation for each of six photometry apertures.  The radii of the apertures, expressed in units of 10$^3$ km, are indicated on the plot.  Lines are added to guide the eye.
\label{bright}}
\end{figure}

\begin{figure}
\epsscale{0.8}
\plotone{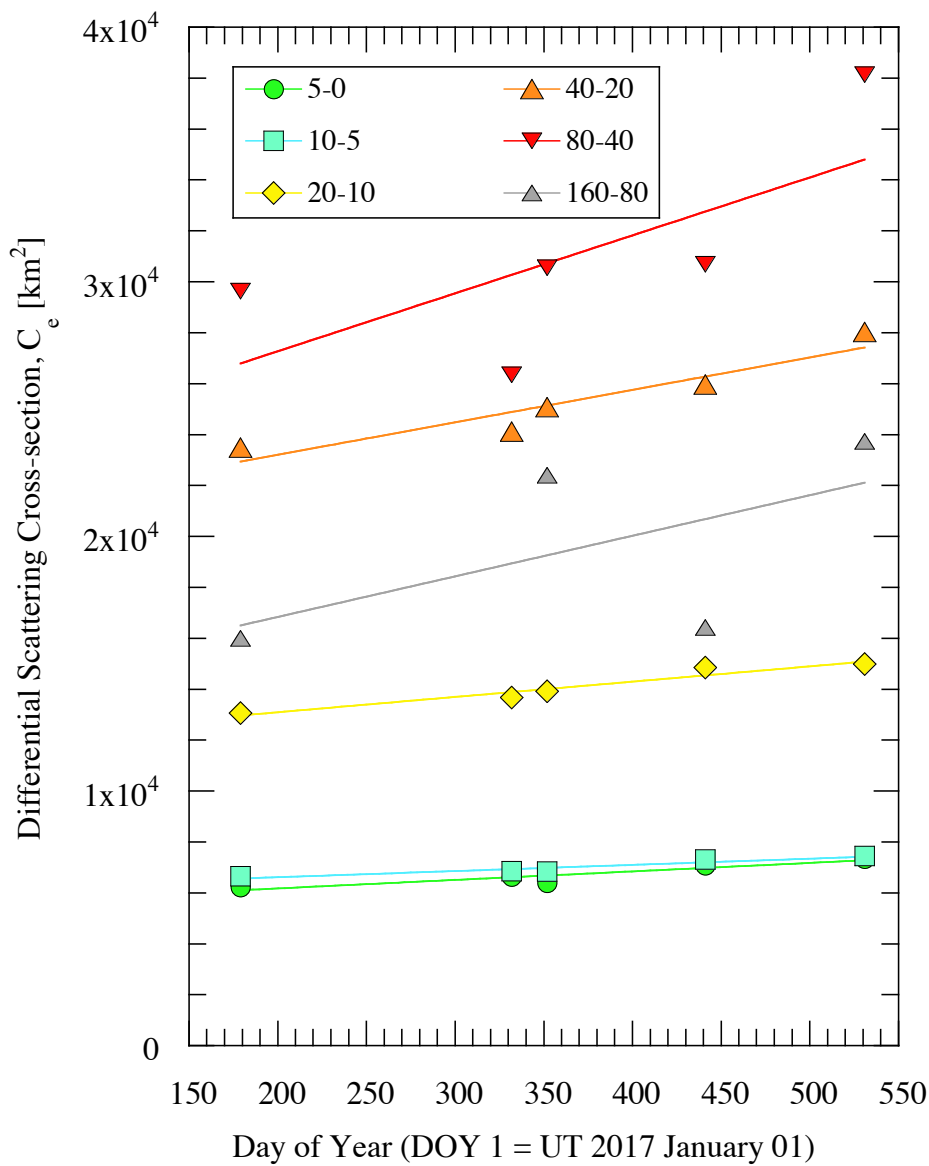}
\caption{Differential scattering cross-section as a function of the date of observation for each of six photometry annuli.  The radii of the apertures, expressed in units of 10$^3$ km, are indicated on the plot (e.g.~20-10 indicates the cross-section in the annulus with inner and outer radii of 10,000 and 20,000 km, respectively).  Lines show least squares fits to the data.  The parameters of these fits are listed in Table (\ref{tablefit}).
\label{bright2}}
\end{figure}

\begin{figure}
\epsscale{1.00}
\plotone{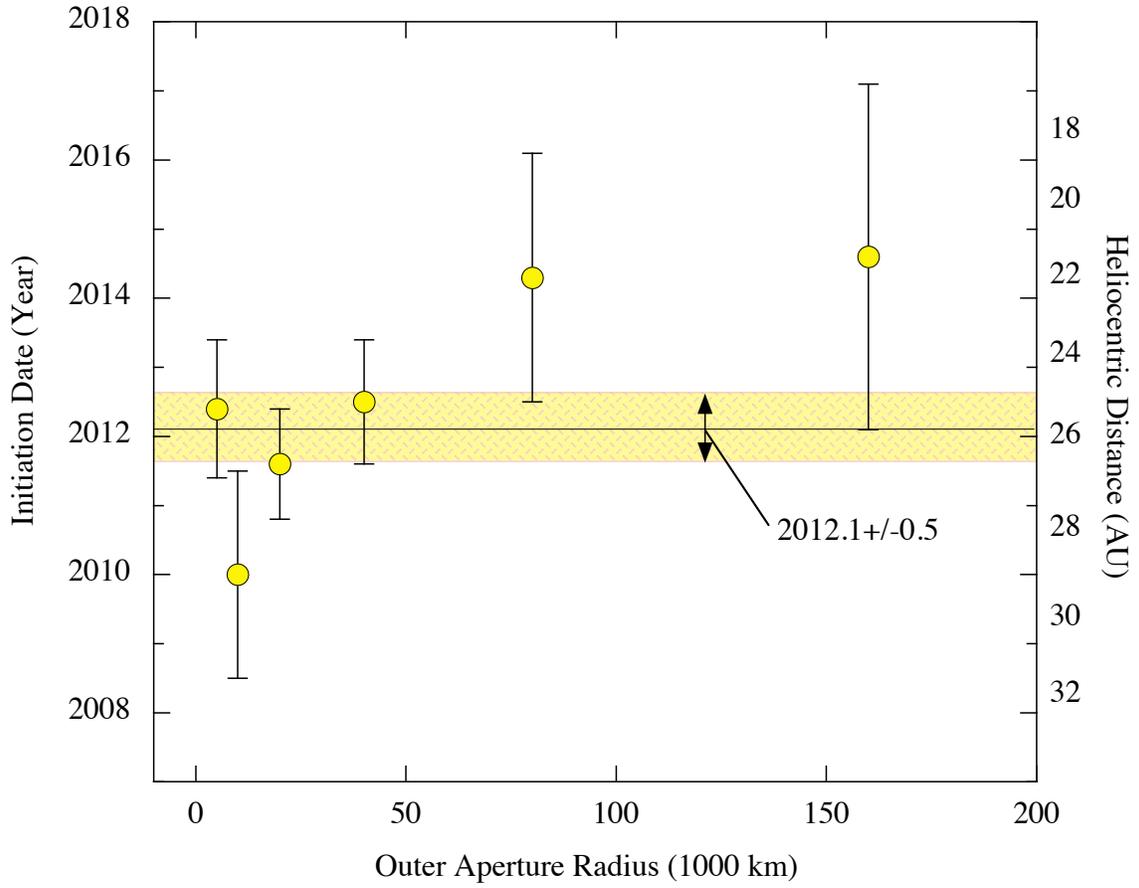}
\caption{Plot of  initiation date vs.~the outer radius of the aperture used to estimate the date.  The weighted mean initiation date is marked together with its $\pm$1$\sigma$ uncertainty, shaded yellow. The right hand axis shows the heliocentric distance. \label{init} }
\end{figure}

The cumulative effective scattering cross-section, $C_e$,  is plotted as a function of the time of observation  in Figure (\ref{bright}), for circular apertures of increasing radius.  All the apertures show a steady brightening of the comet with time which, because we have corrected for geometric effects, must reflect changes in the level of activity of the comet.   Specifically, the brightening reflects the difference between the addition of new material from the nucleus and the loss of older material from the edges of the photometry aperture owing to outward expansion of the coma.  We computed differential cross-sections, $\Delta C_e$, to measure the amount of dust within  each aperture from Table (\ref{photometry}) and show them as a function of time in Figure (\ref{bright2}).  Straight lines in the Figure show  linear least-squares fits  of the form $\Delta C_e = f + gt$, where $t$ is the epoch of observation  and $f$ and $g$ are constants for each annulus.   Best-fit values of $f$ and $g$ are given in Table (\ref{tablefit}), along with the initiation time (i.e.~the time when $\Delta C_e$ = 0) computed from $t_0 = -f/g$.   The latter is expressed both as the number of days prior to UT 2017 January 01 and as decimal year and we further list the heliocentric distance at the initiation time.  As expected from Figures (\ref{bright}) and (\ref{bright2}), uncertainties in the best-fit values of $f$ and $g$, and so of $t_0$, grow with aperture radius because of the effects of noise on the low surface-brightness outer coma.  The annular cross-sections, $\Delta C_e$, and the calculated initiation times, $t_0$, are independent. The weighted mean initiation time is $t_0$ = -1772$\pm$171 DOY, corresponding to year 2012.1$\pm$0.5, at which time the heliocentric distance of K2 was  $r_H = 25.9\pm0.9$ AU (c.f.~Figure \ref{init}).   Clearly, $t_0$ is approximate, since we have assumed without proof that the rate of brightening measured in the period 2017 - 2018 held constant at earlier times. However, we note that our estimated value of $t_0$ precedes the first archival observations that showed  K2 to be active in 2013 May (Jewitt et al.~2017a), by about a year.   We conclude that the data are consistent with activity beginning  $\tau \sim 6.3\pm0.4$ yr ($\sim2\times$10$^8$ s) before the most recent observations (Table \ref{geometry}), when the comet was inbound and $\sim$ 26 AU from the Sun.

We separately fitted the heliocentric variation of the cross-section, represented by a power-law of the form $\Delta C_e \propto r_H^{\gamma}$ finding $\gamma$ = -1.5$\pm$0.6 from a fit to the $\ell$ = 160,000 km aperture photometry (Table \ref{photometry}). The large uncertainty reflects the small fractional change in heliocentric distance over the period of observations.  Within 1$\sigma$, the data are consistent with a production function that varies with the inverse square of $r_H$.

Distant pre-perihelion dust ejection has been inferred in some other comets from the position angles and lengths of their tails.  For example,  Sekanina (1973, 1975)  inferred that dust was ejected from long period comet  C/1954 O2 (formerly 1955 VI (Baade)) starting 500 to 1500 days before perihelion, at heliocentric distances $r_H$ = 6 AU to 12 AU.  Comet C/Bowell (1980b) likewise showed evidence for early (11 or 12 AU), low speed ejection (Sekanina 1982).  Comet C/2010 U3 (Boattini) was discovered at $r_H$ = 18.3 AU inbound, implying coma production at even greater distances.  C/2013 A1 (Siding Spring) ejected 100 $\mu$m particles when beyond Jupiter (Ye and Hui 2014).  Even by the standards of these unusual comets, though, the distant activity in K2 must be regarded as extreme.  

\subsection{Particle Size and Speed}
\label{sizespeed}
A firm lower limit to the dust speed, $U$, can be set by the requirement that dust ejected in 2012 should reach or pass the edge of the 160,000 km radius aperture by 2018.  This limit is $U >$ 0.8 m s$^{-1}$.

A better estimate of the speed can be obtained by considering the motions of the ejected particles under the action of external forces.
 The distance over which radiation pressure can accelerate a particle in time, $t$, is simply
 
 \begin{equation}
 L =  \frac{\beta g_{\odot} }{2 r_H^2} t^2
 \label{radp}
 \end{equation}
 
 \noindent where $g_{\odot}$(1) = 0.006 m s$^{-2}$ is the solar gravity at $r_H$ = 1 AU and $r_H$ is expressed in AU.  The radiation pressure acceleration factor, $\beta$, depends on particle composition and shape but, most importantly, on the particle radius.  To a useful level of approximation for dielectric spheres, we write $\beta = 1/a_{\mu m}$, where $a_{\mu m}$ is the radius expressed in microns.  A 1 $\mu$m sized particle has $\beta \sim$ 1.   For example, 1 $\mu$m particles at $r_H$ = 14 AU would travel $L \sim 5\times10^{11}$ m (3 AU) in  the 6 years since 2012, producing a tail of  angular scale $L/\Delta \sim$ 14\degr~viewed from the distance of K2.  Even given that the angular extent of such a tail would be reduced by a projection factor $\tan(\alpha = 4\degr) \sim$ 0.07, its absence  in Figure (\ref{images}) immediately shows that the particles must be large, with $\beta \ll$ 1 ($a_{\mu m} \gg 1$), as previously noted (Jewitt et al.~2017a).  In fact, if we accept that no tail is evident on angular scales $\theta \sim$ 10\arcsec~(Figure \ref{images}), then Equation (\ref{radp}) gives $\beta < 0.003$ ($a_{\mu m} > 350$), implying the dominance of millimeter-sized particles.

\begin{figure}
\epsscale{0.85}
\plotone{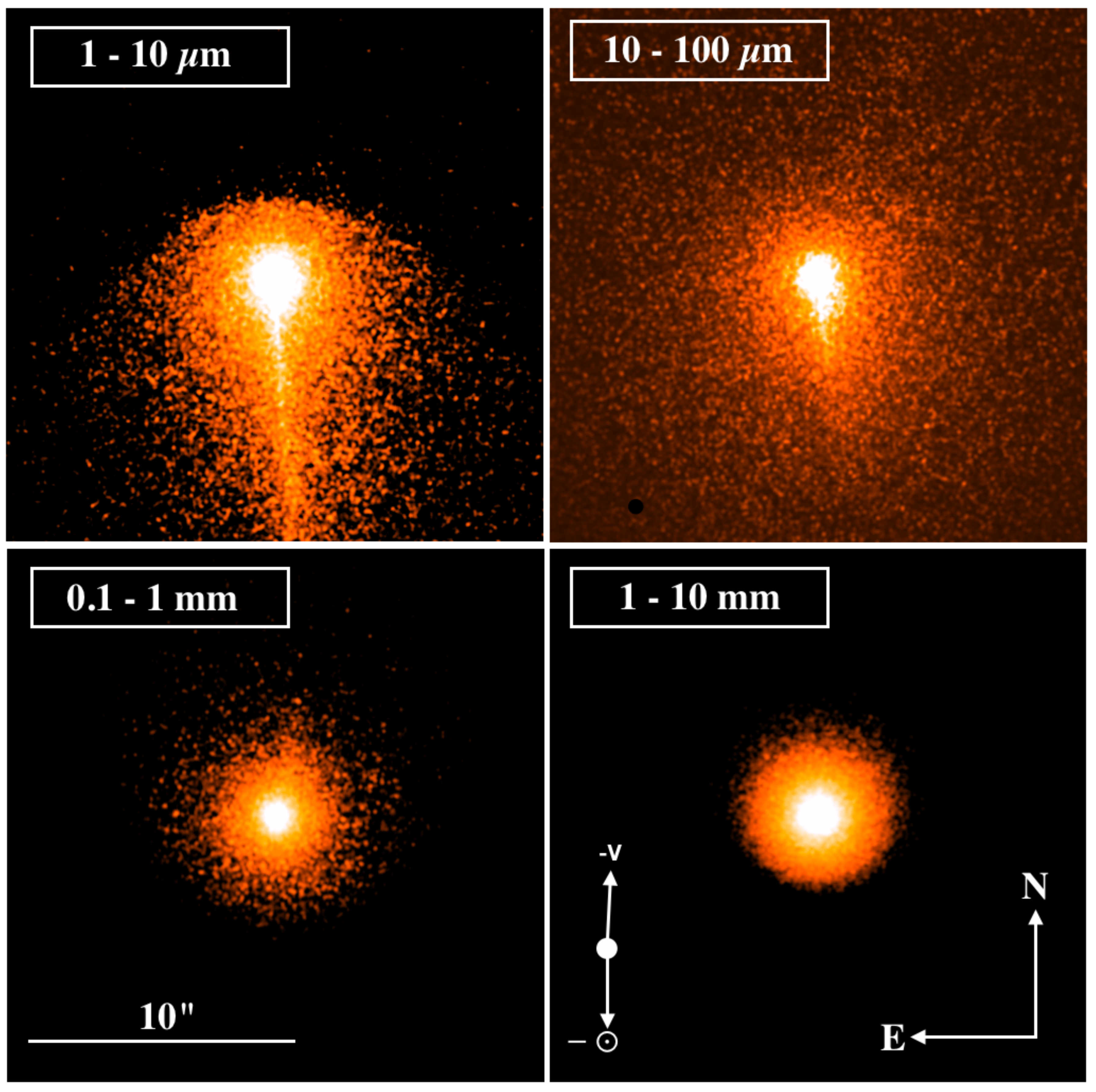}
\caption{Monte-Carlo models for UT 2018 June 15  in which only the sizes of the ejected particles are varied, as marked, to show the influence of radiation pressure.  A scale bar, the cardinal directions and the projected anti-solar and negative velocity vectors are marked.
\label{models}}
\end{figure}

We simulated the appearance of K2 using a Monte-Carlo model to better assess the size of the particles.  In the model, dust particles with a range of sizes are ejected from the sun-facing hemisphere, with activity beginning in 2012.1 as indicated by the data.  The particle sizes occupy a differential power law  distribution with index -3.5 and the ejection velocity, $U$,  is scaled according to $U = U_1 a^{-1/2}$, where $U_1$ is the velocity of a particle having radius $a$ = 1 mm.  The inverse square root relation is expected if gas drag is responsible.   The models were computed for the observation on UT 2018 June 15 because, on this date, the anti-solar and projected orbit directions  are particularly widely separated, clearly exposing the competitive effects of radiation pressure and solar gravity.  We present four examples in Figure (\ref{models}), for particle radii $1 \le a \le 10~\mu$m, $10 \le a \le 100~\mu$m, $100 \le a \le 1000~\mu$m and $1 \le a \le~10$ mm to show the effects of particle size.  The $1 \le a \le 10~\mu$m and $10 \le a \le 100~\mu$m particles are highly responsive to radiation pressure, adopting the familiar umbrella-shaped distribution  with the tail pointed towards the anti-solar direction.  Larger particles, with $100 \le a \le 1000~\mu$m and $1 \le a \le 10$ mm, show a much more nearly isotropic distribution like that in K2 and show, instead, a hint of a (large particle) tail in the direction of the projected orbit.   The simulations confirm that the circular isophotes of K2 reflect particles so large as to be insensitive to solar radiation pressure.  We adopt a nominal minimum particle radius $a$ = 0.1 mm, and we find that   $U_1$ =  4 m s$^{-1}$ (consistent with the lower limit identified above) best represents the scale of the coma in our data.   At this speed, a 1 mm  particle would take $\sim$1 yr to reach the edge of the 160,000 km photometry aperture.   We note that $U$ is comparable to the gravitational escape speed from a non-rotating nucleus of density $\rho$ = 500 kg m$^{-3}$ and radius $r_n$ = 9 km (the upper limit placed by Jewitt et al.~2017a), namely, $V_e$ = 4.8 m s$^{-1}$.  This coincidence suggests that the  particles escaping into the coma are merely the fastest of a larger population of dust grains many of which fell back onto the surface or into sub-orbital trajectories.

\subsection{Dust Production Rates}

The average dust mass production rate is given by

\begin{equation}
\frac{dM}{dt} = \frac{4}{3}\frac{\rho \overline{a} C_e}{\tau_r},
\label{dmbdt}
\end{equation}

\noindent where $\rho$ is the bulk density of the ejected solid material,  $\overline{a}$ is the weighted mean radius of the particles, $C_e$ is the scattering cross-section in a photometric aperture and $\tau_r$ is the time of residence for dust in the aperture.   To obtain a minimum estimate, we take $\overline{a}$ = 0.1 mm, based on the simulations described in section (\ref{sizespeed}). The density of the particles depends on many unknowns, including their composition and their porosity; we assume $\rho = 500$ kg m$^{-3}$.   The cross-section inside the $\ell$ = 160,000 km radius aperture is 119,000 km$^2$ (Table \ref{photometry}) and the residence time is $\tau_r = \ell/U$.  We take $U = 4$ m s$^{-1}$ to find $\tau_r \sim$ 4$\times 10^7$ s.  Then, Equation (\ref{dmbdt}) gives an average mass loss rate $dM/dt =$  200 kg s$^{-1}$ over the period 2012 - 2018.     The corresponding mass in dust is $M \sim$ 10$^{10}$ kg, also good to order-of-magnitude.   Both $M$ and $dM/dt$ should be regarded as order of magnitude estimates, because they are based on several parameters (dust albedo, particle size, density, ejection velocity) that are themselves individually uncertain.

We further represent the nucleus as a sphere of radius $r_n$ and density $\rho$. The coma mass is then equal to the mass contained in a surface shell $\delta r$ thick, where

\begin{equation}
\delta r = \frac{\tau_r}{4\pi\rho r_n^2}\frac{dM}{dt} 
\label{deltar1}
\end{equation}

\noindent Substituting for $dM/dt$ from Equation (\ref{dmbdt}) we obtain 

\begin{equation}
\delta r = \frac{\overline{a} C_e}{3\pi r_n^2}.
\label{deltar2}
\end{equation}

\noindent All three variables in Equation (\ref{deltar2}) are observationally constrained.  Again, taking minimum estimates, $C_e$ = 1.2$\times$10$^5$ km$^2$ (Table \ref{photometry}), $\overline{a} \ge 0.1$ mm (section \ref{sizespeed}) and $r_n \lesssim$ 9 km (Jewitt et al.~2017a), gives  $\delta r \gtrsim$ 1.5 cm.  The material in the coma at the time of observation is equal in mass to a surface shell around the nucleus of K2 only 1.5 cm thick.  Of course, mass loss may be concentrated in a small fraction of the nucleus surface area, leading to much deeper local excavation, as we calculate in the next section.

\subsection{Sublimation}
Temperatures at K2 distances are so low  that only substances more volatile than water ice could sublimate (Jewitt et al.~2017a, Meech et al.~2017).   Carbon monoxide and carbon dioxide are two volatiles that are abundant in comets and which are capable of sublimating in the outer solar system.  To calculate the rates of sublimation of these ices we use the  equilibrium energy balance equation, neglecting conduction.  In this approach, the ice is assumed to be exposed at the cometary surface and so the derived sublimation rates are upper limits to the true values to be expected if, as is likely, the ice is protected by even a thin, refractory layer.  Given the lack of knowledge of the physical properties of K2, however, our simplistic approach provides useful intuition for understanding K2.   

The energy balance equation is

\begin{equation}
\frac{L_{\odot}}{4 \pi r_H^2}(1-A_B)  = \chi\left[ \epsilon \sigma T^4 + L(T) f_s(T)\right].
\label{sublimation}
\end{equation}

\noindent Here, $L_{\odot}$ (W) is the luminosity of the Sun, $r_H$ (m) is the heliocentric distance, $A_B$  is  the Bond albedo of the sublimating material,  $\epsilon$  is the emissivity, $\sigma$ = 5.67$\times$10$^{-8}$ (W m$^{-2}$ K$^{-4}$) is the Stefan-Boltzmann constant and $L(T)$ (J kg$^{-1}$) is the latent heat of sublimation of the relevant ice at temperature, $T$ (K).  The quantity $ f_s$ (kg m$^{-2}$ s$^{-1}$) is the sought-after mass flux of sublimated ice.    The term on the left represents power absorbed from the Sun.  The two terms on the right represent power radiated from the surface into space and power used to sublimate ice.  Dimensionless parameter $1 \le \chi \le 4$ characterizes the way in which incident heat is distributed over the surface. To again focus on the highest possible sublimation rates for a given volatile, we consider sublimation from a flat surface oriented perpendicular to the Sun-comet line, for which  $\chi$ = 1, which approximates conditions found at the sub-solar point on a non-rotating nucleus.  To solve Equation (\ref{sublimation}) we assumed $L_{\odot}$ = 4$\times$10$^{26}$ W, $\epsilon$ = 0.9, $A_B$ = 0.04 and used ice thermodynamic parameters for CO and CO$_2$ from Brown and Ziegler (1980) and Washburn (1926).   

The solutions to Equation (\ref{sublimation}) give $f_s \sim 2\times10^{-5}$ kg m$^{-2}$ s$^{-1}$ for CO and $f_s \sim 6\times10^{-6}$ kg m$^{-2}$ s$^{-1}$ for CO$_2$, both for $r_H$ = 16 AU, and these are peak values because we have assumed that the ice is located at the subsolar point on K2, this being the hottest spot.  In order to supply a  dust production rate, $dM/dt \sim$ 200 kg s$^{-1}$, would require a sublimating area 

\begin{equation}
A_s = \frac{dM/dt}{f_{dg} f_s}
\label{dg}
\end{equation}

\noindent where $f_{dg}$ is the ratio of the dust to gas mass production rates.  This quantity is unmeasured in distant comets, but in Jupiter family comets close to the Sun it is generally $>$ 1 (e.g.~Reach et al.~2000, Fulle et al.~2016).  We conservatively adopt $f_{dg}$ = 5 and find, from Equation (\ref{dg}), sublimation areas $A_s = 2\times10^6$ m$^2$ (2 km$^2$) for CO and $A_s = 6\times10^6$ m$^2$ (6 km$^2$) for CO$_2$.  Both values are small compared to the surface area of a 9 km radius nucleus ($4\pi r_n^2 \sim 10^3$ km$^2$) showing that even a small patch of exposed ice can account for the observed activity.  The surface erosion rate due to sublimation  is $d r_n/dt = f_s/\rho$ (m s$^{-1}$).  With density $\rho$ = 500 kg m$^{-3}$, we find $d r_n/dt \sim 1.2$ m yr$^{-1}$ for CO and $\sim$ 0.4 m yr$^{-1}$ for CO$_2$.  In the 6 years since activity began, sublimation would have eroded the nucleus locally by a very modest $\delta r \lesssim$ 10 m, perhaps creating pit and cliff morphology like that observed on the nucleus of 67P/Churyumov-Gerasimenko.

\subsection{The Cohesion Bottleneck}
The escaping gas must have enough momentum to drag dust particles from the nucleus.  A minimum condition is obtained by simply equating the gas drag force with the weight of the dust particle, as was done first by Whipple (1950).  From this condition we calculate $a_c$, the radius of the largest grain  which gas can eject against nucleus gravity,  from

\begin{equation}
a_c(r_H)= \frac{9 C_D V_g(r_H) f_s(r_H)}{16 \pi G \rho \rho_n  r_n}.
\label{ac}
\end{equation}

Quantity $C_D$ is a dimensionless drag coefficient representing the efficiency with which gas momentum is transferred to the grain, $V_g$ is the gas outflow speed, $\rho$ and $\rho_n$ are the densities of the grains and nucleus, respectively, and  $r_n$ is the nucleus radius.  We assume $C_D$ = 1, $\rho_n$ = $\rho$ = 500 kg m$^{-3}$, $V_g$ = 100 m s$^{-1}$ and $r_n$ = 9 km.  The numerical constants in Equation (\ref{ac}) are specific to the assumption of spherical particles and should not be regarded as important.  What matters is that $a_c$ is most sensitive to $f_s$, evaluated from Equation (\ref{sublimation}).

Equation (\ref{ac}) is a necessary but not sufficient condition for grain ejection because of the existence of inter-particle cohesive forces, as clearly described by Gundlach et al. (2015).  Experiments show that the cohesive strength of micron-sized aggregates is about 150 N m$^{-2}$, and that the strength scales inversely with the size (Sanchez and Scheeres 2014).  Accordingly,  the cohesive strength of an assemblage of particles of radius $a$ may be written as

\begin{equation}
S = S_0 / a
\label{ss14}
\end{equation}

\noindent where $S_0 \sim 1.5\times 10^{-4}$ (N m$^{-1}$) and $a$ is expressed in meters (c.f.~Sanchez and Scheeres 2014).  Note that a slightly different size dependence ($S \propto a^{-2/3}$) is preferred by Gundlach et al.~(2018).  To estimate the effects of particle cohesion on the escape of particles, we compare the sublimation gas pressure, $P_s = f_s V_{g}$, where $V_{g}$ is the thermal speed in the gas at the equilibrium sublimation temperature, $T$, with the cohesive strength, $S$.  We set $P_s = S$ to find the critical radius, $a_s$, above which gas drag forces exceed the inter-particle cohesion, finding

\begin{equation}
a_s(r_H) = \frac{S_0}{C_D f_s(r_H) V_g(r_H)}.
\label{as}
\end{equation}

\noindent Again, we use Equation (\ref{sublimation}) to solve for $f_s$ and $V_g$ as functions of $r_H$ and Equation (\ref{as}) to calculate the critical  radius, $a_s$.   Note that $a_s \propto f_s^{-1}$, opposite to the dependence of $a_c$.

Equations (\ref{ac}) and (\ref{as}) are plotted for CO sublimation in Figure (\ref{co_allowed}) and for CO$_2$ sublimation in Figure (\ref{co2_allowed}).   We refer to the point where the curves cross as the ``cohesion bottleneck''.  Setting $a_c$ = $a_s$, we solve Equations (\ref{ac}) and (\ref{as}) for

\begin{equation}
f_s = \left(\frac{16 \pi G \rho \rho_n r_n S_0}{9 C_D^2 V_g^2}\right)^{1/2}
\end{equation}

\noindent and, by substitution, obtain $f_s = 10^{-4}$ kg m$^{-2}$ s$^{-1}$ at the bottleneck.  By solution of Equation (\ref{sublimation}), this $f_s$ is reached by CO at $r_H \sim$ 6.6 AU and CO$_2$ at $r_H \sim$ 4.5 AU.  At distances smaller than the crossing point, there exists a range of particles that are both small enough to be lifted against gravity and large enough to be lifted against cohesive forces.  At distances larger than the crossing point, particles which are small enough to be lifted against  nucleus gravity are trapped by inter-particle cohesion.  Our observations of millimeter-sized particles in distant comet K2  fall entirely in this latter regime, where inter-particle cohesive forces prohibit grain ejection.

\begin{figure}
\epsscale{0.8}
\plotone{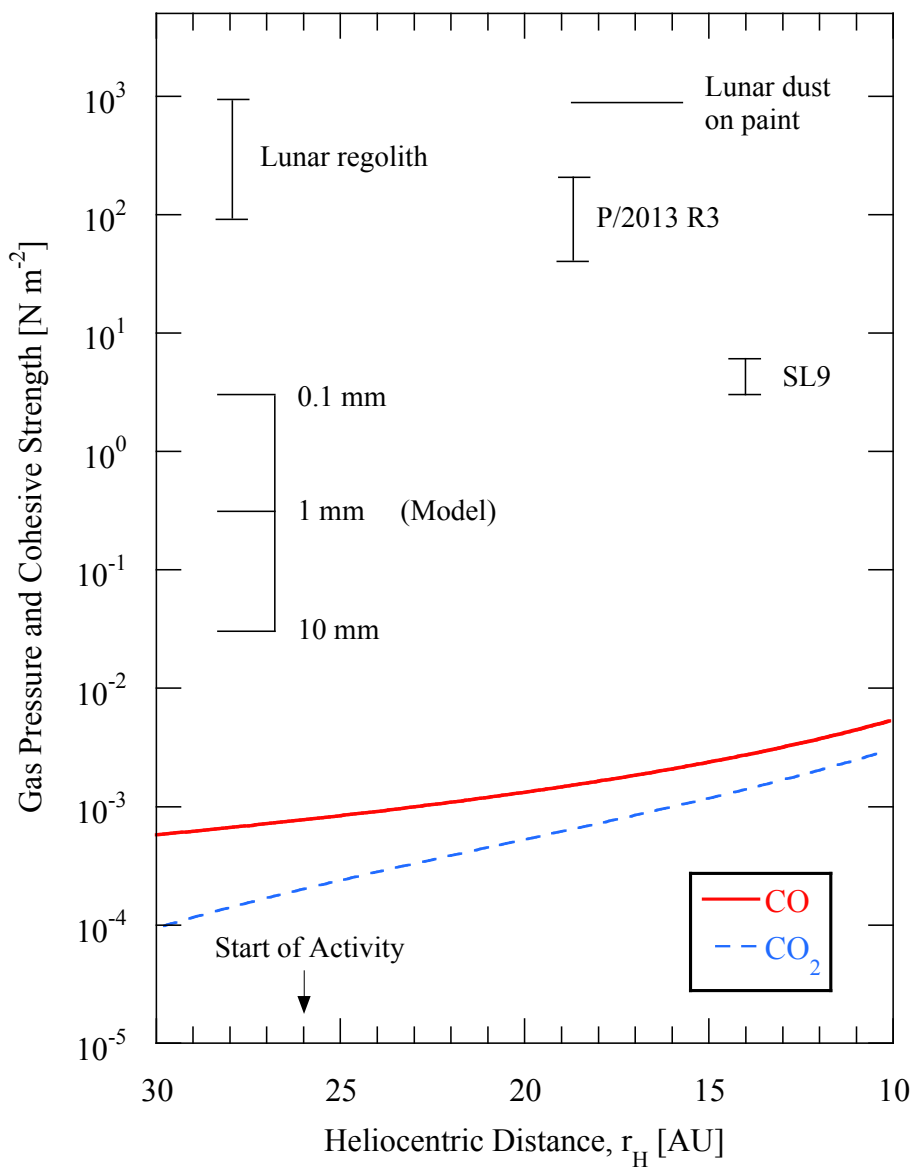}
\caption{Gas pressure produced by the equilibrium sublimation of CO (solid red curve) and CO$_2$ (dashed blue curve) compared with cohesion estimates from several sources: Lunar regolith (Mitchell et al.~1974), Lunar dust adhering to paint (Scott and Zuckerman 1971), disrupted active asteroid P/2013 R3 (Jewitt et al.~2014, 2017, Hirabayashi et al.~2014), tidally split P/Shoemaker-Levy 9 (SL9) (Asphaug and Benz 1996) and a cohesion model for three particle sizes (Sanchez and Scheeres 2014).   \label{pressure} }
\end{figure}

To emphasize that sublimation  is unable to expel particles from the nucleus against inter-particle cohesion, we plot solutions for the sublimation gas pressure from Equation (\ref{sublimation}) as a function of heliocentric distance in Figure (\ref{pressure}).  The Figure shows a slightly higher sublimation pressure for CO (solid red curve) than for CO$_2$ (dashed blue curve), as expected from the greater volatility of CO.  Also shown in Figure (\ref{pressure}) are two estimates of the cohesive strength inferred for lunar regolith dust (Mitchell et al.~1974, Scott and Zuckerman 1971), for the fragmented active asteroid P/2013 R3 (Jewitt et al.~2014, 2017b, Hirabayashi et al.~2014) and for split comet P/Shoemaker-Levy 9 (Asphaug and Benz 1996), together with strength estimates from Equation (\ref{ss14}) for three particle sizes.  The sublimation gas pressures, even of CO, are 2 to 3 orders of magnitude too small to overcome the interparticle cohesion in the Sanchez and Scheeres (2014) model, and 4 to 6 orders of magnitude smaller than cohesive strengths measured in the regolith of the Moon, or inferred from break-up of P/2013 R3 and SL9.  

We attempted to increase the calculated sublimation pressure by changing parameters in the sublimation Equation (\ref{sublimation}), without success.  This is because, for volatile substances, the sublimation term already takes up a large fraction of the absorbed power, leaving little room for the radiation term to make a difference.  The calculated pressures apply to free sublimation into a vacuum, matching the conditions at the surface of a comet nucleus. Any increase in the temperature above the equilibrium value results in an increase in the sublimation rate (i.e.~the energy is used to break inter-molecular bonds in the ice), which quickly depresses the temperature back to the equlibrium value.  Even setting $\epsilon$ = 0 in Equation (\ref{sublimation}) (i.e.~forcing all the absorbed energy into the sublimation term)  increases the calculated $f_s(CO)$ by only a few percent.  The increase is larger (factor of 2.6) for $f_s(CO_2)$  but this still falls far short of the orders of magnitude strength gap in Figure (\ref{pressure}).     In this sense, the mechanism for the ejection of dust from K2 is a mystery.  We expect that activity beyond the cohesion bottleneck is a general property of all distant comets.

\subsection{Overcoming the Bottleneck}
The failure of sublimation gas drag forces to overcome the cohesion bottleneck  motivates brief consideration of  other processes  that might expel dust from comets at large heliocentric distances.  

\begin{figure}
\epsscale{0.8}
\plotone{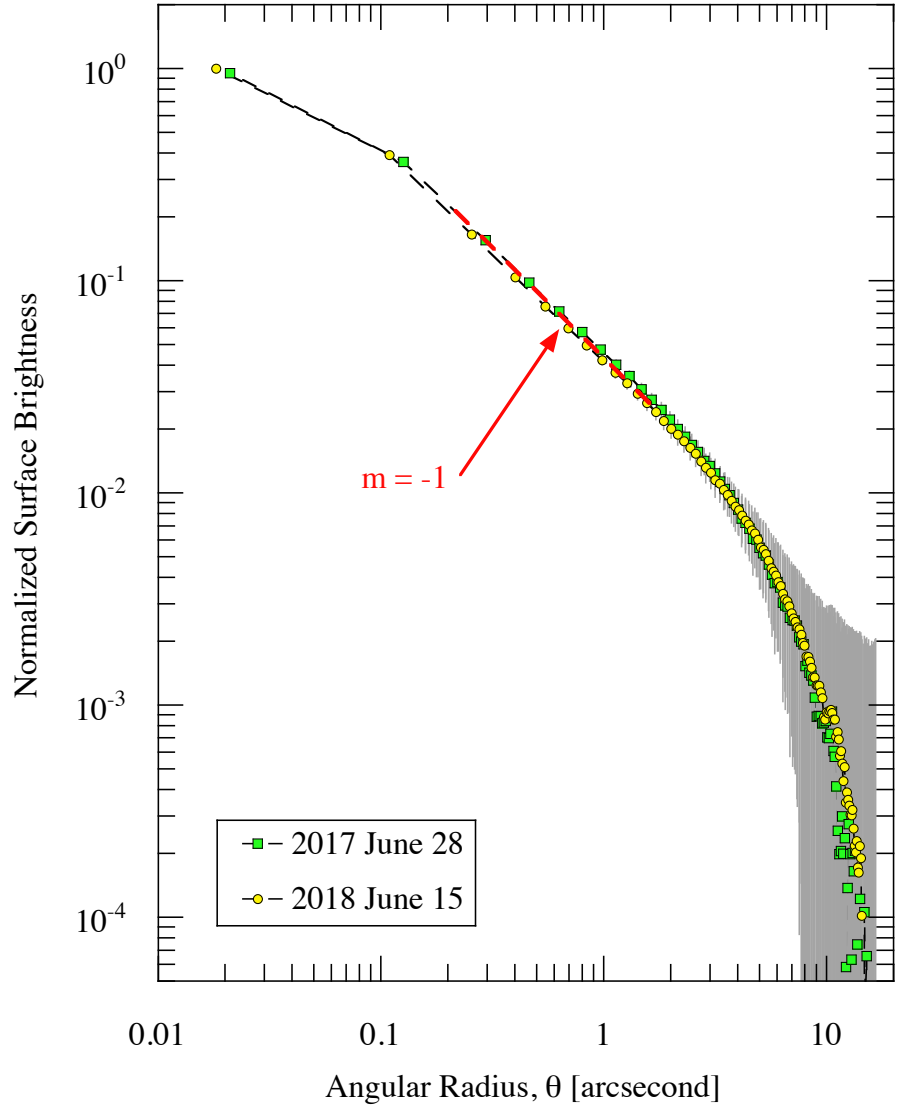}
\caption{Comparison of surface brightness profiles measured in 2017 June (green squares) and 2018 June (yellow circles).  Profile uncertainties resulting from the sky background are shown by the shaded grey region.  The red line shows a logarithmic gradient $m$ = -1, expected of steady-state coma production.
\label{SB}}
\end{figure}

Prolonged exposure  of cometary ices to the ionizing cosmic ray flux is expected to cause molecular damage and the build-up  of  unstable radicals in a surface layer perhaps  $\sim$1 m thick (Cooper et al.~2003).  Explosive recombination of these radicals upon warming by the Sun has been suggested as a source of activity in distant comets (Donn and Urey 1956) and could conceivably generate instantaneous gas pressure large enough to overcome the cohesion bottleneck.  Unfortunately, the steady-state coma surface brightness gradient (Figure \ref{SB}) and the slow, monotonic brightening of K2 (Figure \ref{bright2}) are inconsistent with the burst-like activity that would result from recombination. Consequently, we do not consider explosive radical recombination to be plausible.  

Impacts are occasionally suggested as causes of  activity in distant comets (e.g.~Cikota et al.~2018).  However,  even in the ecliptic the probability of impact is small and K2 (with inclination $i$ = 87.5\degr) is located far above the mid-plane of the solar system, leading to a negligible probability of collision.  Furthermore, dust produced impulsively by an impact would  violate the photometric and surface brightness evidence to the effect that the coma is long-lived (Figure \ref{images}), and produced in steady-state  (Figure \ref{SB}).  Neither would impacts selectively eject millimeter-sized debris, as observed.  We confidently rule out the possibility of impact.

At $\sim$25 AU, even at the maximum possible  (subsolar) temperature,  $T_{SS} \sim$  79 K, the timescale  for the crystallization of ice is comparable to the age of the solar system (Kouchi and Yamamoto 1995).  Therefore, even if amorphous ice exists on K2, its crystallization can also be rejected as a mechanism for the observed early activity (c.f.~Jewitt et al.~2017a).  Crystallization cannot become a significant source of gas until the comet has reached the $r_H \sim$ 10 to 12 AU region (Guilbert-Lepoutre 2012), if at all.  

\begin{figure}
\epsscale{0.95}
\plotone{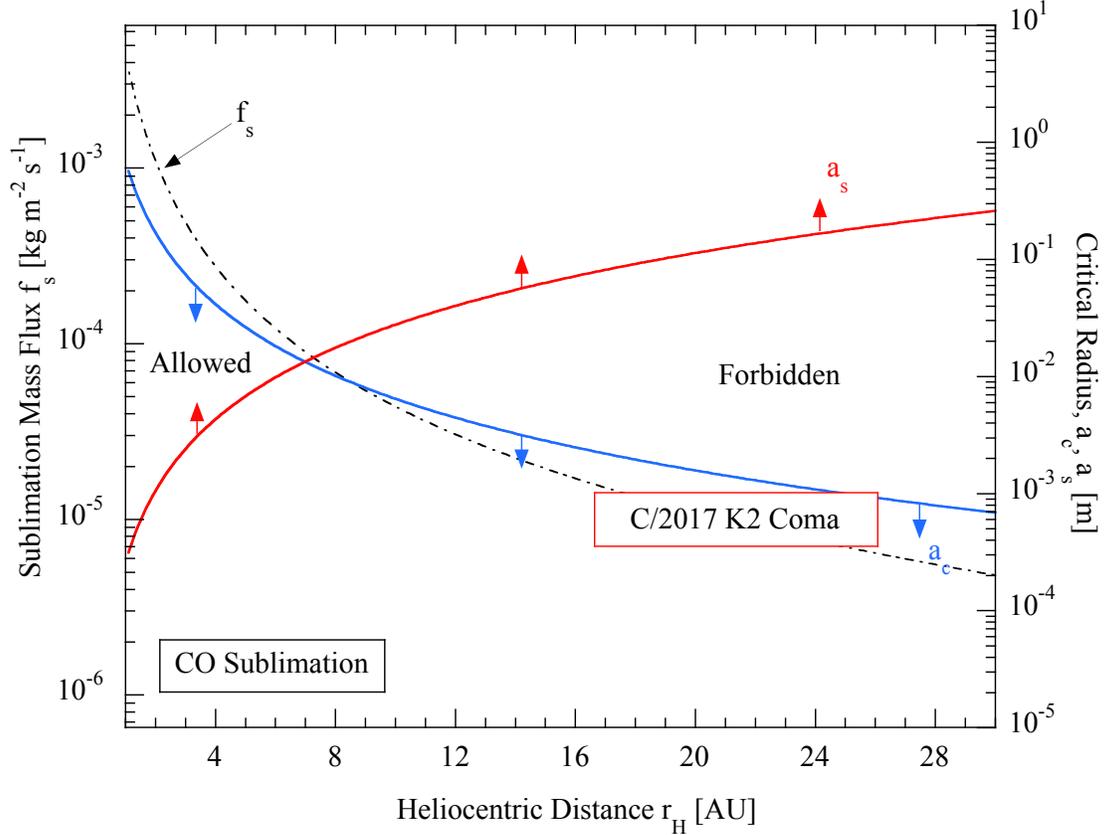}
\caption{The equilibrium carbon monoxide specific sublimation rate (Equation \ref{sublimation}) is  shown by the dashed black curve (left-hand axis). The blue curve shows the maximum ejectable dust grain radius against nucleus gravity as a function of heliocentric distance (Equation \ref{ac}). The red curve shows the  minimum ejectable grain radius for overcoming inter-particle cohesion (Equation \ref{as}).   To be ejected, a particle must plot above the red cohesion curve and below the blue gravity curve, which is impossible at the distance of K2.  Critical radii are labelled on the right-hand axis.  \label{co_allowed} }
\end{figure}

\begin{figure}
\epsscale{0.95}
\plotone{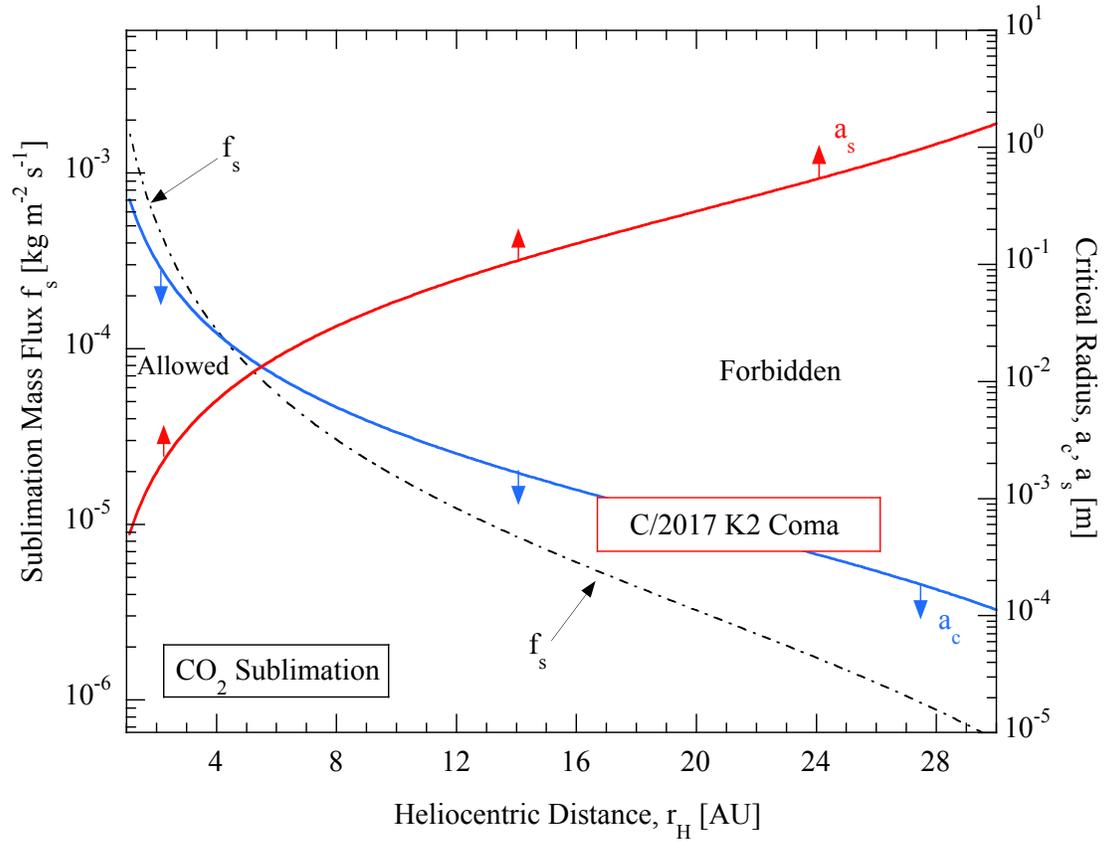}
\caption{Same as Figure (\ref{co_allowed}) but for carbon dioxide sublimation.    \label{co2_allowed} }
\end{figure}

A related but different process known as ``annealing'' occurs when molecular bonds in amorphous ice locally rearrange, releasing trapped gases  prior to  larger-scale crystallization.  We find only very limited experimental data  about annealing in the literature (e.g.~Ninio-Greenberg et al.~2017, Yokochi 2018), but these data suggest that it is unlikely to be a significant process in K2.  For example, the experimentally determined fluxes of  nitrogen exuded by annealing at 70 K are $\sim$10$^{18}$ m$^2$ s$^{-1}$ (Ninio-Greenberg et al.~2017), corresponding to mass flux $f_s \sim 5\times 10^{-8}$ kg m$^{-2}$ s$^{-1}$.  This is two to three orders of magnitude smaller than the CO sublimation mass flux at K2's distance (Figure \ref{co_allowed}).  We tentatively conclude that annealing offers no solution to overcoming the cohesion bottleneck.

Thermal fracture is another process that might operate at the very low temperatures present on K2, with the temperature contrast arising between the subsolar and night-side temperatures.  This contrast is a function of both the thermal diffusivity and the rotation period, neither of which is known.  Experiments show that ice at 77 K  has a Young's modulus, $Y = 1.8\times 10^{10}$ N m$^{-2}$, and is very brittle, with fracture beginning at stresses $S \sim 3\times 10^5$ N m$^{-2}$ and failure (described by the authors as ``explosive'') at $S \sim 4\times 10^7$ N m$^{-2}$ (Parameswaran and Jones 1975).  Ignoring Poisson's ratio, the strain induced by temperature difference $\delta T$ is $\sigma \sim \varepsilon \delta T$, where $\varepsilon \sim 7\times 10^{-6}$ K$^{-1}$ is the expansivity at 77 K.  Setting $Y = S/\sigma$, we solve to find 

\begin{equation}
\delta T = \frac{S}{\varepsilon Y}
\end{equation}

\noindent and, by substitution we find $\delta T$ = 3 K for first fracture and 300 K for failure.  Of course, $\delta T$ = 300 K can never be reached on a comet because the ice would sublimate away.  Furthermore, the mechanical properties sampled in the laboratory are those of crystalline ice and amorphous ice in comets might have different values. Still, $\delta T$ = 3 K is a very small temperature threshold for the onset of fracture and it is plausible that, upon approaching the Sun, diurnal temperature fluctuations could reach the level needed to fracture ice.  Fracture is attractive because stored energy from thermal stresses  would be able to break the cohesion bottleneck, leaving particles free to be expelled against nucleus gravity alone.

Dielectric surfaces exposed to sunlight  develop a positive charge as a result of the loss of photoelectrons.  Evidence that particles can be levitated by near-surface electric fields comes from ``horizon glow'' observations on the Moon (Criswell and De 1977) and, with less certainty, from ponded particulates that have migrated across the surfaces of asteroids into local gravitational potential minima (Robinson et al.~2001).  Unfortunately, the mechanism by which particles might be electrostatically detached from the lunar regolith is not understood, although it is highly relevant to understanding the cohesion bottleneck, in general,  and the case of K2, in particular.  The surface potential raised by photoionization is only $\sim$5 V and, across a meter-thick Debye layer, the resulting electric field is only $\sim$5 V m$^{-1}$. The resulting repulsive force is negligible compared to the weight and to cohesive forces.  

Theoretical consideration of  micro-scale super-charging suggests that, depending on the conductivity and other properties, huge electric fields (e.g.~10$^5$ V m$^{-1}$) might be generated between near-surface particles in a regolith (Zimmerman et al.~2016).  In recent experiments this super-charging has been shown to launch 5 $\mu$m sized grains at $\sim$1 m s$^{-1}$, almost sufficient to explain the lunar horizon glow (Wang et al.~2016).  On a kilometer scale cometary nucleus, these launch velocities approach the gravitational escape speed and direct escape to the coma would be possible, even without the aid of post-detachment gas drag.  

While promising, it is not obvious that an effect capable of detaching grains at the surface of the Moon can operate at the edge of the planetary region.  For example, the ionizing flux varies as $r_H^{-2}$, meaning that the charging currents at 26 AU will be $\sim$700 times smaller than at 1 AU, and the charging timescales 700 times longer.  Charging times of 1 day at 1 AU correspond to $\sim$2 years at 26 AU.  Long charging times will elevate the importance of leakage currents between nearby grains owing to finite electrical conductivity, so making large potential differences and strong electric fields harder to attain.  Moreover, the effect described in Wang et al.~is  still too weak to launch millimeter sized grains at 4 m s$^{-1}$ and so it does not directly account for the coma of K2.  Nevertheless, we remain cautiously optimistic that  electrostatic supercharging can break the cohesion bottleneck and we encourage more work on this process.

Thermal fracture and electrostatic supercharging operate independently of gas drag and it must be considered  possible that the coma of K2 is  produced without the aid of gas.  More likely, gas drag, fracture and/or electrostatic supercharging operate in tandem to cause cometary activity at ultra-large heliocentric distances.   As K2 approaches the Sun, the role of non-thermal effects will diminish relative to the rising sublimation mass flux, culminating in the entry of the comet into the ``allowed'' region of Figures (\ref{co_allowed}) and (\ref{co2_allowed}).

\subsection{The Future}
The slow approach of K2 towards perihelion on 2022 December 21, at $r_H = 1.810$ AU, offers the opportunity to examine previously unstudied  changes in an inbound Oort cloud comet.   We anticipate a change in the style of mass loss with the onset of crystallization at about $r_H \lesssim$ 12 AU (starting early 2019) and another change as the comet approaches the orbit of Jupiter (5 AU, in late 2021), where water ice can begin to sublimate.   Water ice grains in the coma will themselves sublimate at this distance, leading to a decrease in the scattering cross-section even as nucleus activity rises.  Thus, it is impossible to predict the brightness evolution of the comet, with brightening caused by increased heating of the nucleus by the Sun, perhaps augmented by exothermic crystallization and the release of trapped gas, but countered by a fading due to the loss of ice from dust grains in the coma.

\clearpage

\section{SUMMARY}
New measurements of inbound, long-period comet C/2017 K2 at heliocentric distances $r_H$ = 15.9 AU to 13.8 AU, give the following main results:

\begin{enumerate}

\item Extrapolation of the photometry shows that mass loss from K2 started  in 2012.1$\pm$0.5 (when $r_H$ = 25.9$\pm$0.9 AU).

\item Dust is released continuously and in steady-state, as shown by the surface brightness gradient of the coma, $m = -1.010\pm0.004$.  The gradient is inconsistent with a coma produced by impulsive activity such as might result from an outburst, allowing us to rule out  gas release from the explosive recombination of radicals, and interplanetary impacts, as plausible sources of distant activity in C/2017 K2.  
The comet is also far too cold for crystallization of amorphous ice to drive the coma. 

\item We find that the optically dominant particles are large (radius $\gtrsim$ 0.1 mm), rendering them insensitive to deflection by solar radiation pressure and explaining why the isophotes of the coma are persistently circular.  Dust ejection speeds, $\sim$4 m s$^{-1}$,  are small compared to the thermal speed in gas ($\sim$10$^2$ m s$^{-1}$) but comparable to the $\sim$5 m s$^{-1}$ gravitational escape speed of a few kilometer radius nucleus.   The order-of-magnitude dust mass loss rate is $\sim 200$ kg s$^{-1}$.

\item Drag forces from sublimating supervolatiles (e.g.~CO, but to a lesser extent, CO$_2$)  exceed the gravitational attraction to the nucleus for millimeter-sized and smaller particles. However, these drag forces are orders of magnitude too small to overcome inter-particle cohesion, defining the ``cohesion bottleneck'' and raising the puzzle of how dust ejection at large distances is possible.   

\item We suggest  thermal fracture and electrostatic super-charging in the cometary regolith (the latter recently postulated to explain horizon-glow observations on the Moon) as two previously neglected processes having the potential to overcome the cohesion bottleneck.

\end{enumerate}

\acknowledgments
We thank Bin Yang for reading the paper.  Based on observations made under GO 15409 and 15423 with the NASA/ESA Hubble Space Telescope, obtained at the Space Telescope Science Institute,  operated by the Association of Universities for Research in Astronomy, Inc., under NASA contract NAS 5-26555.   JA appreciates funding from European Research Council Starting Grant No. 757390.



{\it Facilities:}  \facility{HST}.

\clearpage


\clearpage

\begin{deluxetable}{lccrrrccccr}
\tablecaption{Observing Geometry 
\label{geometry}}
\tablewidth{0pt}
\tablehead{\colhead{UT Date \& Time} & \colhead{DOY\tablenotemark{a}} & \colhead{$r_H$\tablenotemark{b}} & \colhead{$\Delta$\tablenotemark{c}}  & \colhead{$\alpha$\tablenotemark{d}} & \colhead{$\theta_{- \odot}$\tablenotemark{e}} & \colhead{$\theta_{-V}$\tablenotemark{f}} & \colhead{$\delta_{\oplus}$\tablenotemark{g}}   }

\startdata

2017 Jun 28 20:09 - 20:52\tablenotemark{h} & 179 & 15.869 & 15.811 & 3.67 &  166.3  & 357.2  & 0.59 \\

2017 Nov 28 17:08 - 17:52 			& 332 & 14.979 & 15.133 & 3.71&  17.1 & 358.3 & 1.38\\

2017 Dec 18 22:30 - 23:13 			& 352 & 14.859 & 15.010 & 3.73 &  358.4 & 357.3  & 0.08 \\
2018 Mar 17 09:28 - 10:34 			& 441 & 14.331 &14.328  & 3.98 &  275.6  & 353.3  & -3.97 \\
2018 Jun 15  15:40 - 16:19 			& 531 & 13.784 & 13.668 & 4.21 &  181.3 & 356.3 & -0.28 \\

\enddata


\tablenotetext{a}{Day of Year, DOY = 1 on UT 2017 January 01}
\tablenotetext{b}{Heliocentric distance, in AU}
\tablenotetext{c}{Geocentric distance, in AU}
\tablenotetext{d}{Phase angle, in degrees}
\tablenotetext{e}{Position angle of projected anti-solar direction, in degrees}
\tablenotetext{f}{Position angle of negative projected orbit vector, in degrees}
\tablenotetext{g}{Angle from orbital plane, in degrees}
\tablenotetext{h}{Observations from GO 14939, described in Jewitt et al.~(2017)}

\end{deluxetable}

\clearpage

\begin{deluxetable}{lcccccc}
\tabletypesize{\scriptsize}

\tablecaption{HST Fixed-Aperture Photometry\tablenotemark{a} 
\label{photometry}}
\tablewidth{0pt}

\tablehead{ \colhead{UT Date} & $\ell/10^3$ = 5 & 10  & 20 & 40 & 80 & 160}
\startdata
2017 Jun 28 &  	21.59/9.45/6.22 & 20.80/8.66/12.9 & 20.04/7.90/25.9	& 19.34/7.20/49.4 & 18.83/6.69/79.1 & 18.63/6.49/95.1  \\
2017 Nov 28                            & 21.31/9.38/6.64         & 20.54/8.61/13.5   & 19.78/7.85/27.2   & 19.09/7.16/51.3   &  18.64/6.71/77.6 & --- \\

2017 Dec 18                            & 21.31/9.42/6.40         & 20.52/8.63/13.2   & 19.74/7.85/27.2    & 19.03/7.14/52.2   & 18.53/6.64/82.8 & 18.27/6.38/105.2 \\

2018 Mar 17   & 21.03/9.31/7.08 & 20.26/8.54/14.4 & 19.49/7.77/29.2 &  18.80/7.08/55.2  & 18.32/6.60/85.9 & 18.13/6.41/102.9\\
2018 Jun 15   & 20.81/9.27/7.34 & 20.05/8.51/14.8 & 19.29/7.75/29.8 & 18.57/7.03/57.8   & 18.02/6.48/95.9  & 17.78/6.24/119.0 \\

\enddata

\tablenotetext{a}{For each date and aperture radius, $\ell$ (measured in units of 10$^3$ km at the comet), the Table lists the apparent magnitude, V, the absolute magnitude, H, and the scattering crossection, $C_e$ (in units of 10$^3$ km$^2$), in the order V/H/$C_e$. $C_e$ is computed from H using Equation (\ref{area}).}

\end{deluxetable}

\clearpage

\begin{deluxetable}{cccrllcc}

\tablecaption{Fixed-Aperture Photometry vs.~Time 
\label{tablefit}}
\tablewidth{0pt}

\tablehead{\colhead{Radius\tablenotemark{a}} & \colhead{$f$\tablenotemark{b} [km$^2$]} & \colhead{$g$\tablenotemark{b} [km$^2$ day$^{-1}$]} & \colhead{$t_0$(days)\tablenotemark{c}}  & \colhead{$\tau_0$(yr)\tablenotemark{d}} & \colhead{$r_H$(AU)\tablenotemark{e}}}
\startdata
$C_e(5)$ - $C_e(0)$ & 5515$\pm$280 & 3.3$\pm$0.7 & -1671$\pm$364 & 2012.4$\pm$1.0 & 25.5$\pm$1.6\\

$C_e(10)$ - $C_e(5)$ & 6134$\pm$184 & 2.4$\pm$0.5 & -2556$\pm$538 & 2010.0$\pm$1.5 & 29.3$\pm$2.4  \\

$C_e(20)$ - $C_e(10)$ & 11896$\pm$344 & 6.0$\pm$0.9 & -1982$\pm$303 &  2011.6$\pm$0.8 & 26.8$\pm$1.2 \\

$C_e(40)$ - $C_e(20)$ & 20662$\pm$984 & 12.7$\pm$2.6 & -1626$\pm$342 & 2012.5$\pm$0.9 & 25.2$\pm$1.7\\

$C_e(80)$ - $C_e(40)$ & 22727$\pm$5279 & 22.7$\pm$13.7 & -1001$\pm$647 & 2014.3$\pm$1.8 & 22.0$\pm$3.2\\

$C_e(160)$ - $C_e(80)$ & 13651$\pm$6003 & 15.9$\pm$15.1 & -859$\pm$899 & 2014.6$\pm$2.5 & 21.6$\pm$4.3\\
\hline
Weighted Mean &  & &  -1772$\pm$171  & 2012.1$\pm$0.5 & 25.9$\pm$0.9\\

\enddata

\tablenotetext{a}{Projected radius of the photometry annulus, in units of 10$^3$ km}
\tablenotetext{b}{Coefficients $f$ and $g$ of the least-squares fit}
\tablenotetext{c}{The initiation time, computed from $t_0 = f/g$, expressed as the number of days prior to UT 2017 January 01}
\tablenotetext{d}{The initiation time, expressed in decimal year}
\tablenotetext{e}{Heliocentric distance, AU}

\end{deluxetable}

\clearpage

\clearpage

\clearpage

\clearpage

\clearpage

\clearpage

\clearpage

\clearpage 

\clearpage

\clearpage

\clearpage

\clearpage


\clearpage

\begin{thebibliography}{}




\bibitem[Asphaug \& Benz(1996)]{1996Icar..121..225A} Asphaug, E., \& Benz, W.\ 1996, \icarus, 121, 225 


\bibitem[Brown\& Ziegler(1980)]{1980BrownAdvCryoEng}  Brown, G. and Ziegler W. (1980).  Adv.~Cryog.~Eng.~25, 662-670.

\bibitem[Cikota et al.(2018)]{2018MNRAS.475.2512C} Cikota, S., Fern{\'a}ndez-Valenzuela, E., Ortiz, J.~L., et al.\ 2018, \mnras, 475, 2512 

\bibitem[Cooper et al.(2003)]{2003EM&P...92..261C} Cooper, J.~F., Christian, E.~R., Richardson, J.~D., \& Wang, C.\ 2003, Earth Moon and Planets, 92, 261 

\bibitem[Criswell \& de(1977)]{1977JGR....82..999C} Criswell, D.~R., \& de, B.~R.\ 1977, \jgr, 82, 999 

\bibitem[de la Fuente Marcos \& de la Fuente Marcos(2018)]{2018RNAAS...2b..10D} de la Fuente Marcos, R., \& de la Fuente Marcos, C.\ 2018, Research Notes of the American Astronomical Society, 2, 10 

\bibitem[Donn \& Urey(1956)]{1956ApJ...123..339D} Donn, B., \& Urey, H.~C.\ 1956, \apj, 123, 339 

\bibitem[Fulle et al.(2016)]{2016MNRAS.462S...2F} Fulle, M., Altobelli, N., Buratti, B., et al.\ 2016, \mnras, 462, S2 

\bibitem[Greenberg et al.(2017)]{2017MNRAS} Greenberg, A. N., Laufer, D., and Bar-Nun, A. \ 2017, \mnras, 469, S517

\bibitem[Guilbert-Lepoutre(2012)]{2012AJ....144...97G} Guilbert-Lepoutre, A.\ 2012, \aj, 144, 97 

\bibitem[Gulkis et al.(2015)]{2015Sci...347a0709G} Gulkis, S., Allen, M., von Allmen, P., et al.\ 2015, Science, 347, aaa0709 


\bibitem[Gundlach et al.(2015)]{2015A&A...583A..12G} Gundlach, B., Blum, J., Keller, H.~U., \& Skorov, Y.~V.\ 2015, \aap, 583, A12 

\bibitem[Gundlach et al.(2018)]{2018MNRAS.479.1273G} Gundlach, B., Schmidt, K.~P., Kreuzig, C., et al.\ 2018, \mnras, 479, 1273 

\bibitem[Hirabayashi et al.(2014)]{2014ApJ...789L..12H} Hirabayashi, M., Scheeres, D.~J., S{\'a}nchez, D.~P., \& Gabriel, T.\ 2014, \apjl, 789, L12 

\bibitem[Hui et al.(2018)]{2018AJ....155...25H} Hui, M.-T., Jewitt, D., \& Clark, D.\ 2018, \aj, 155, 25 

\bibitem[Jewitt \& Meech(1987)]{1987ApJ...317..992J} Jewitt, D.~C., \& Meech, K.~J.\ 1987, \apj, 317, 992 

\bibitem[Jewitt et al.(2014)]{2014ApJ...784L...8J} Jewitt, D., Agarwal, J., Li, J., et al.\ 2014, \apjl, 784, L8 

\bibitem[Jewitt et al.(2017)]{2017ApJ...847L..19J} Jewitt, D., Hui, M.-T., Mutchler, M., et al.\ 2017a, \apjl, 847, L19 

\bibitem[Jewitt et al.(2017)]{2017AJ....153..223J} Jewitt, D., Agarwal, J., Li, J., et al.\ 2017b, \aj, 153, 223 

\bibitem[Kouchi \& Yamamoto(1995)]{1995PCGC...153} Kouchi, A., and Yamamoto, T. \ 1995, Prog. Crystal Growth and Charact., 30, 83


\bibitem[Kr{\'o}likowska \& Dybczy{\'n}ski(2018)]{2018A&A...615A.170K} Kr{\'o}likowska, M., \& Dybczy{\'n}ski, P.~A.\ 2018, \aap, 615, A170 
%
%

\bibitem[Meech et al.(2017)]{2017ApJ...849L...8M} Meech, K.~J., Kleyna, J.~T., Hainaut, O., et al.\ 2017, \apjl, 849, L8 

\bibitem[Mitchell et al.(1972)]{1972LPI.....3..545M} Mitchell, J.~K., Scott, R.~F., Houston, W.~N., et al.\ 1972, Lunar and Planetary Science Conference, 3, 545 

\bibitem[Ninio-Greenberg et al.(2017)]{2017MNRAS.469S.517N} Ninio-Greenberg, A., Laufer, D., \& Bar-Nun, A.\ 2017, \mnras, 469, S517 

\bibitem[Parameswaran et al.(1975)]{1975...JGlac} Parameswaran, V., and Jones, S.  \ 1975, J. Glaciology, 14, 305

\bibitem[Reach et al.(2000)]{2000Icar..148...80R} Reach, W.~T., Sykes, M.~V., Lien, D., \& Davies, J.~K.\ 2000, \icarus, 148, 80 


\bibitem[Robinson et al.(2001)]{2001Natur.413..396R} Robinson, M.~S., Thomas, P.~C., Veverka, J., Murchie, S., \& Carcich, B.\ 2001, \nat, 413, 396 

\bibitem[S{\'a}nchez \& Scheeres(2014)]{2014M&PS...49..788S} S{\'a}nchez, P., \& Scheeres, D.~J.\ 2014, Meteoritics and Planetary Science, 49, 788 


\bibitem[Scheeres \& S{\'a}nchez(2018)]{2018PEPS....5...25S} Scheeres, D.~J., \& S{\'a}nchez, P.\ 2018, Progress in Earth and Planetary Science, 5, 25 

\bibitem[Scott \& Zuckerman(1971)]{1971LPSC....2.2743S} Scott, R.~F., \& Zuckerman, K.~A.\ 1971, Lunar and Planetary Science Conference Proceedings, 2, 2743 

\bibitem[Sekanina(1973)]{1973ApL....14..175S} Sekanina, Z.\ 1973, \aplett, 14, 175 

\bibitem[Sekanina(1975)]{1975Icar...25..218S} Sekanina, Z.\ 1975, \icarus, 25, 218 

\bibitem[Sekanina(1982)]{1982AJ.....87..161S} Sekanina, Z.\ 1982, \aj, 87, 161 

\bibitem[Wang et al.(2016)]{2016GeoRL..43.6103W} Wang, X., Schwan, J., Hsu, H.-W., Gr{\"u}n, E., \& Hor{\'a}nyi, M.\ 2016, \grl, 43, 6103 

\bibitem[Washburn(1926)]{1926washburn} Washburn, E.\ 1926, International Critical Tables of Numerical data, Physics, Chemistry and Technology, Vol. 3 (New York: McGraw-Hill).

\bibitem[Whipple(1950)]{1950ApJ...111..375W} Whipple, F.~L.\ 1950, \apj, 111, 375 

\bibitem[Ye \& Hui(2014)]{2014ApJ...787..115Y} Ye, Q.-Z., \& Hui, M.-T.\ 2014, \apj, 787, 115 


\bibitem[Yokochi(2018)]{2018LPI....49.2957Y} Yokochi, R.\ 2018, Lunar and Planetary Science Conference, 49, 2957 

\bibitem[Zimmerman et al.(2016)]{2016JGRE..121.2150Z} Zimmerman, M.~I., Farrell, W.~M., Hartzell, C.~M., et al.\ 2016, Journal of Geophysical Research (Planets), 121, 2150 

\end{thebibliography}
\end{document}